\documentclass{icrc29}
\usepackage{graphicx,amssymb,amsmath,times}
\begin{document}
%Title of paper
\title[Observations of AGN ...]{Observations of AGN with the First VERITAS Telescope}
\author[P.Cogan et al.] {P.Cogan$^a$ for The VERITAS Collaboration$^b$ \\
        (a) Dept. Experimental Physics, University College Dublin, Belfield, Dublin 4, Ireland \\ 
        (b) For full author list, see J. Holder's paper ``Status and Performance of the First VERITAS Telescope'' from these proceedings
        }
\presenter{Presenter: P. Cogan (peter@ferdia.ucd.ie),ire-cogan-P-abs1-og23-poster}

\maketitle

\begin{abstract}
The first VERITAS (Very Energetic Radiation Imaging Telescope
Array System) telescope has been in operation at the basecamp of the
Whipple Observatory since January 2005. Here we present initial
observations of AGN made using this telescope. Although this is
engineering data, significant detections of Markarian 421 and Markarian 501 have
been achieved.

\end{abstract}

\section{Introduction}
During the first 5 months of 2005, the first VERITAS Telescope was
operated as a stand-alone Imaging Atmospheric Cherenkov 
Telescope (IACT) at the basecamp of the Whipple Observatory in southern Arizona. The Crab
Nebula, which is regarded as the standard candle of TeV gamma-ray
astronomy, was a priority target during this period. The detection of
the Crab Nebula was announced with the First Light declaration on
February 1st 2005 and is reported elsewhere \cite{holder}.

Following this detection, observations of the Crab Nebula continued in
conjunction with observations of established TeV emitters such as
Markarian 421 and Markarian 501.

%, 1ES 1426+428 and 1ES
%1959+650. Observations of very high energy (VHE) candidates 1ES
%1011+496, 1ES 1741+196 and RGB J1725+118 also commenced at this time.

In this paper the scientific motivations for observing Active Galactic
Nuclei with VERITAS will be outlined, the telescope will be briefly
described and the preliminary analysis of the AGN data will be
presented.

%Significances and fluxes are reported on Markarian 421 and
%Markarian 501.% and upper limits are presented on the other targets.

\section{Scientific Motivation}

%AGN (Active Galactic Nuclei) are galaxies whose central cores greatly
%outshine the rest of the host galaxy. The central engine of an AGN is
%believed to be powered by a super massive black hole which draws
%surrounding matter into a circulating accretion disk. Infalling matter
%loses angular momentum through viscous or turbulent forces and powers
%the accretion disk whose thermal emission peaks at UV
%wavelengths. Beyond the accretion disk lies a dusty region, commonly
%regarded as torus shaped, which absorbs visible and UV light along
%some lines of sight. Jets of energetic particles stream from the poles
%of the torus, perpendicularly away from the plane of the host
%galaxy. These collimated relativistic plasma outflows harbour some of
%the most energetic processes in the universe and exhibit emission from
%radio to gamma ray. This unifying model of AGNs \cite{urryandpadovani}
%implies that the observed TeV emission from an AGN is a strong
%function of the observation angle in relation to the relativistic jet,
%with the highest energy emission beamed along the direction of the
%jet. The unifying model describes blazars as those AGN whose
%collimated plasma jets are closely aligned with our galaxy. Such a
%chance orientation provides unique opportunities to study these jets
%and to understand he underlying emission mechanisms in AGN.

The non-thermal emission of Active Galactic Nuclei (AGN) is well
characterised by a $\nu F_ \nu$ plot \cite{urryandpadovani}. This
Spectral Energy Distribution (SED) shows two broad peaks; the lower
ranges from the infrared to the x-ray whereas the higher peaks at
gamma-ray energies. The lower energy peak is generally attributed to
synchrotron emission by highly relativistic electrons within plasma
jets which propagate perpendicularly away from the plane of the host
galaxy \cite{bloommarscher}. There is less certainty regarding the
origin of the higher energy peak and several plausible models
exist. In the Inverse Compton (IC) model, high-energy emission is
produced through inverse-Compton scattering of the relativistic
electrons with either the locally produced synchrotron photons
(Synchrotron Self Compton model) or with photons from the external
environment (External Compton model). In a hadronic model, protons are
shock accelerated in the jet to extremely high energies. These protons
can collide with target nuclei in the jet and produce neutral pions
which decay to high energy photons \cite{mannheim}. Alternatively high
energy photons can be produced via proton-synchrotron emission. Only
through observations of flux and spectral variability over a broad
range of wavelengths can the emission mechanisms be determined.

An understanding of the extra-galactic background light (EBL) is
necessary for studies of star and galaxy formation. However, measurement
of the EBL in the optical to far infra red is inhibited by foreground
radiation such as that due to diffuse dust in the Milky Way Galaxy. Very high
energy (VHE) photons are absorbed by the EBL via pair production. For
those AGN which have been detected, an attenuation of the spectral index
with increasing red shift has been observed\cite{schroedter}. This can
be interpreted as scattering of VHE photons off the EBL at higher energies. If the
intrinsic AGN spectrum is known, information about the shape of the
EBL spectrum can be extracted and would shed new light on the
evolution of the universe. Thus far there are 6 confirmed AGN which
have contributed to an understanding of the EBL. In order to further constrain this background, a large
catalog of AGN at various redshifts must be established. This will be
a primary goal of VERITAS.

%\cite{daviescotton}
\section{VERITAS Telescope 1}

VERITAS Telescope 1 was operated as a stand-alone IACT instrument from
January through June 2005 \cite{holder} following autumn upgrades to
the VERITAS prototype \cite{cogan}. The telescope is temporarily
located at the basecamp of the Whipple Observatory. This is not at an
optimal observing elevation and is not well shielded from background light pollution. The telescope uses the Davies Cotton
\cite{daviescotton} reflector design and has a focal length of 12m,
making it an f/1 system for a 12m aperture. The telescope comprises
340 mirrors giving a surface area of 100 m$^2$. The telescope's camera
has 499 PMTs with a field of view of 3.5$^\circ$, however light cones have not yet been installed.

%VERITAS Telescope 1 utilises a flexible three-level trigger system which allows
%full-array, sub-array and individual telescope operation. Trigger
%level one consists of constant-fraction discriminators (CFDs)
%\cite{hall} which discriminate against night-sky background. The level
%two trigger is a topological hardware trigger which can discriminate
%between compact Cherenkov events and random night-sky or
%afterpulse-induced events \cite{bradburyetal}. Trigger level three is
%an array trigger. For VERITAS Telescope 1 only the first two trigger
%levels were required.

VERITAS Telescope 1 utilises a two-level trigger system. Trigger level
one consists of constant-fraction discriminators (CFDs) which
discriminate against night-sky background \cite{hall}. The level-two trigger is a
topological hardware trigger which can discriminate between compact
Cherenkov events and random night-sky or afterpulse-induced events \cite{bradburyetal}.

The PMT signals are digitised by a custom-designed 500 MHz
flash-analogue-to-digital converter (FADC) system
\cite{buckleyetal}. The data acquisition has 50 FADC boards,
containing 10 channels each, located in four VME crates. For every
event, the FADCs provide a pulse profile for each pixel. As well as
improving the signal to noise ratio, reducing the need for delay
cables and reducing deadtime, the FADCs may provide a new technique
for discriminating against hadronic events by using the pulse profiles
to study the temporal evolution of the extensive air shower
\cite{holder}.

\section{Observations/Analysis}

All observations were taken in PAIR mode where the source is observed
for 28 minutes (ON), followed two minutes later by a similar exposure
(OFF) with an offset of 30 minutes in Right Ascension (RA) in order to
assess the background. For this dataset, the analysis was tested and
optimised using observations of the Crab Nebula \cite{holder}.

% Whereas Markarian 421, Markarian 501,
%1ES 1426+428 and 1ES 1956+650 are established TeV sources, 1ES
%1011+496, 1ES 1741+196 and RGBJ 1725+118 have never been detected at
%TeV energies. They were selected as candidates for VHE emission due to
%their RA, red shift and emission at other wavelengths.

%In order to calibrate the telescope, the camera is illuminated nightly
%with a nitrogen dye laser. These five minute laser runs are used to
%produce gain and timing corrections. The telescope is artifically
%triggered at 3 Hz in order to generate pedestals. The RMS spread of the
%pedestals is a measure of the noise in the pixel and is used in
%cleaning the image. An attenuated level 2 trigger pulse is piped into a spare
%channel on each of the four FADC crates to ensure read-out synchronicity.

\begin{figure}[h]
\begin{center}
\includegraphics*[width=0.3\textwidth,angle=270,clip]{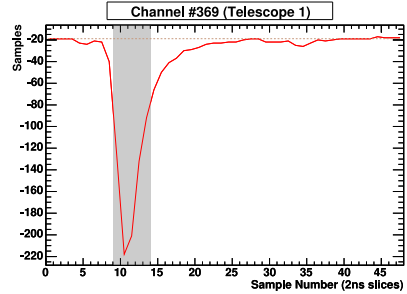} 
\caption{\label{trace} FADC trace from a laser pulse. The optimised summation window is indicated by the shaded area and the pedestal is indicated by the dashed line. Sampling is performed at intervals of two nanoseconds. }
\end{center}
\end{figure}

\begin{figure}[h]
\begin{center}
\includegraphics*[width=0.3\textwidth,angle=270,clip]{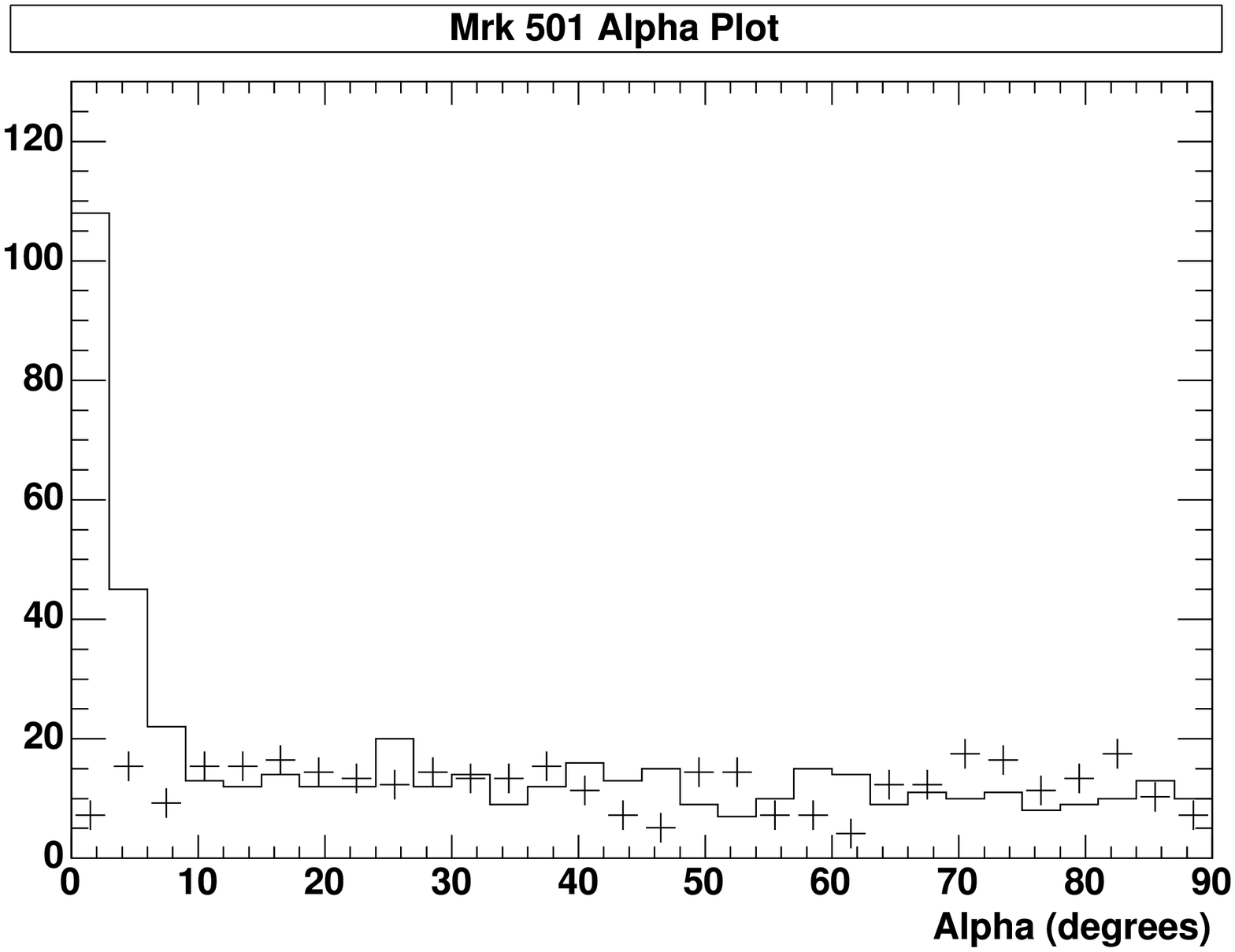} 
\includegraphics*[width=0.3\textwidth,angle=270,clip]{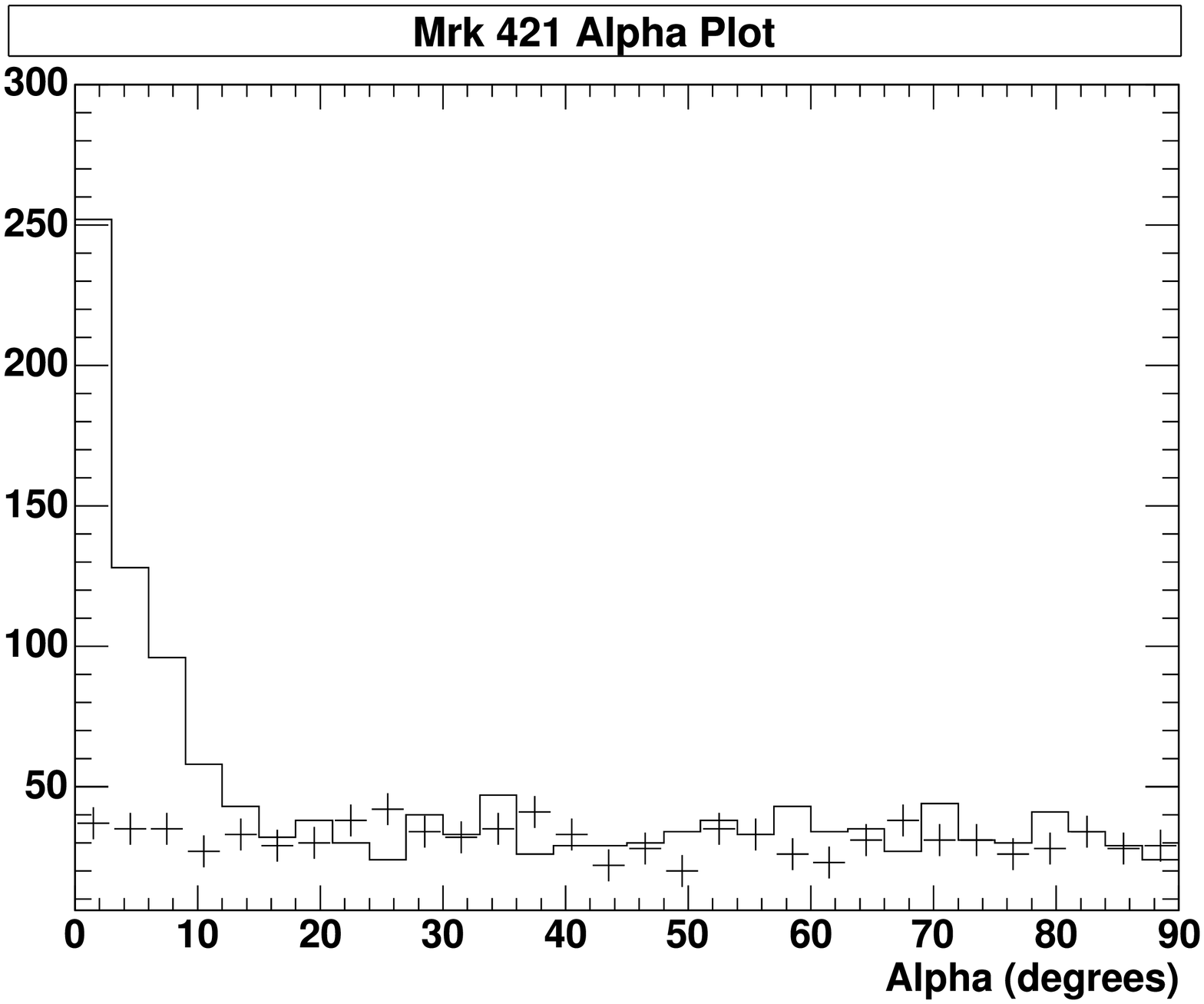}
\caption{\label{alphaplots} \emph{Left:} Alpha distribution for Markarian
501. \emph{Right:} Alpha distribution for Markarian 421. The ON source
distribution is represented by the solid line, with the OFF source
distribution represented as crosses.}
\end{center}
\end{figure}

\begin{figure}[h]
\begin{center}
\includegraphics*[width=0.35\textwidth,angle=270,clip]{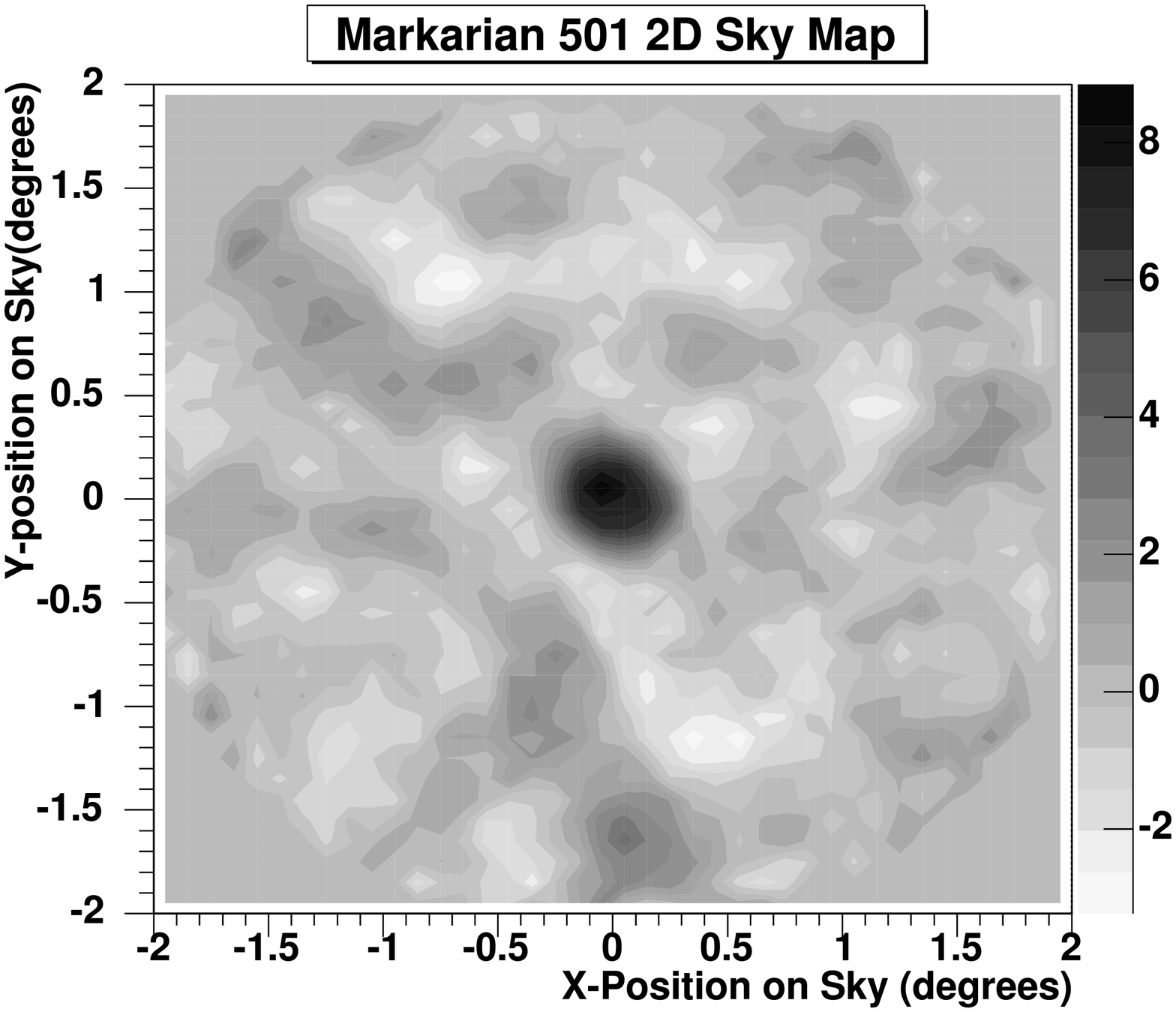} 
\includegraphics*[width=0.35\textwidth,angle=270,clip]{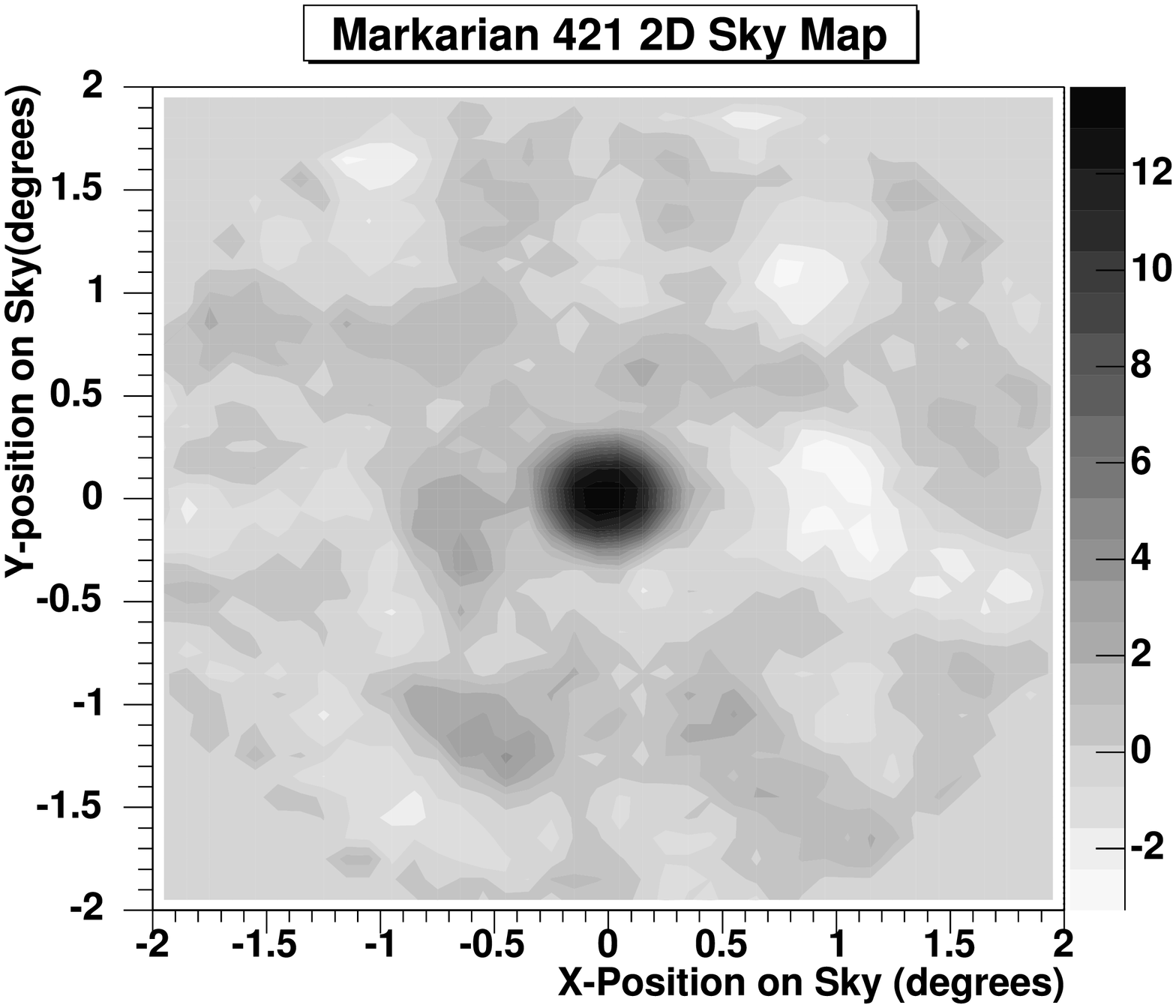}
\caption{\label{skymaps} \emph{Left:} Two-dimensional significance map for
Markarian 501. \emph{Right:} Two-dimensional significance map for Markarian
421. The colour scale refers to significance levels (note different scales). Adjacent bins are not independent.}
\end{center}
\end{figure}

\begin{figure}[h]
\begin{center}
\includegraphics*[width=0.3\textwidth,angle=270,clip]{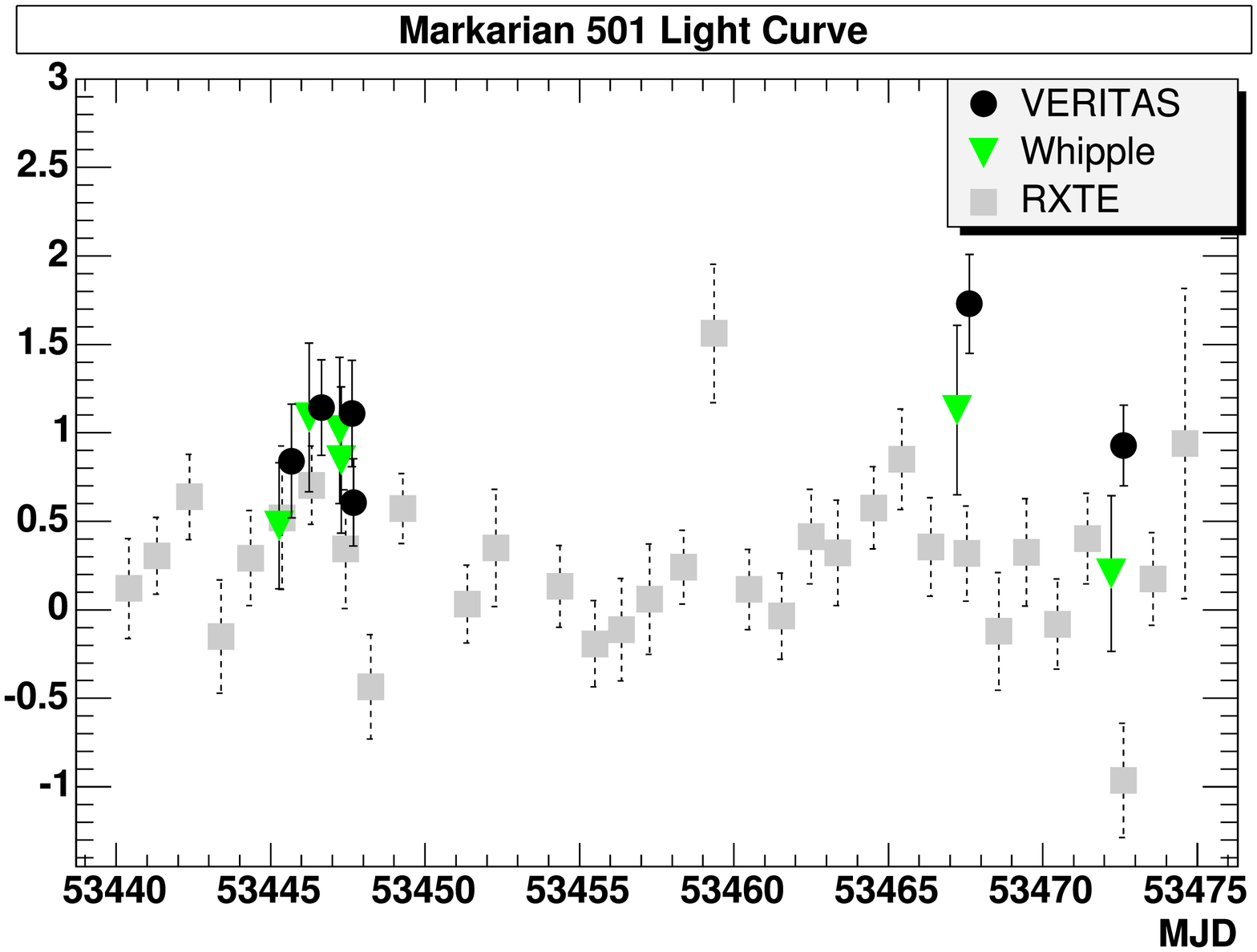} 
\includegraphics*[width=0.3\textwidth,angle=270,clip]{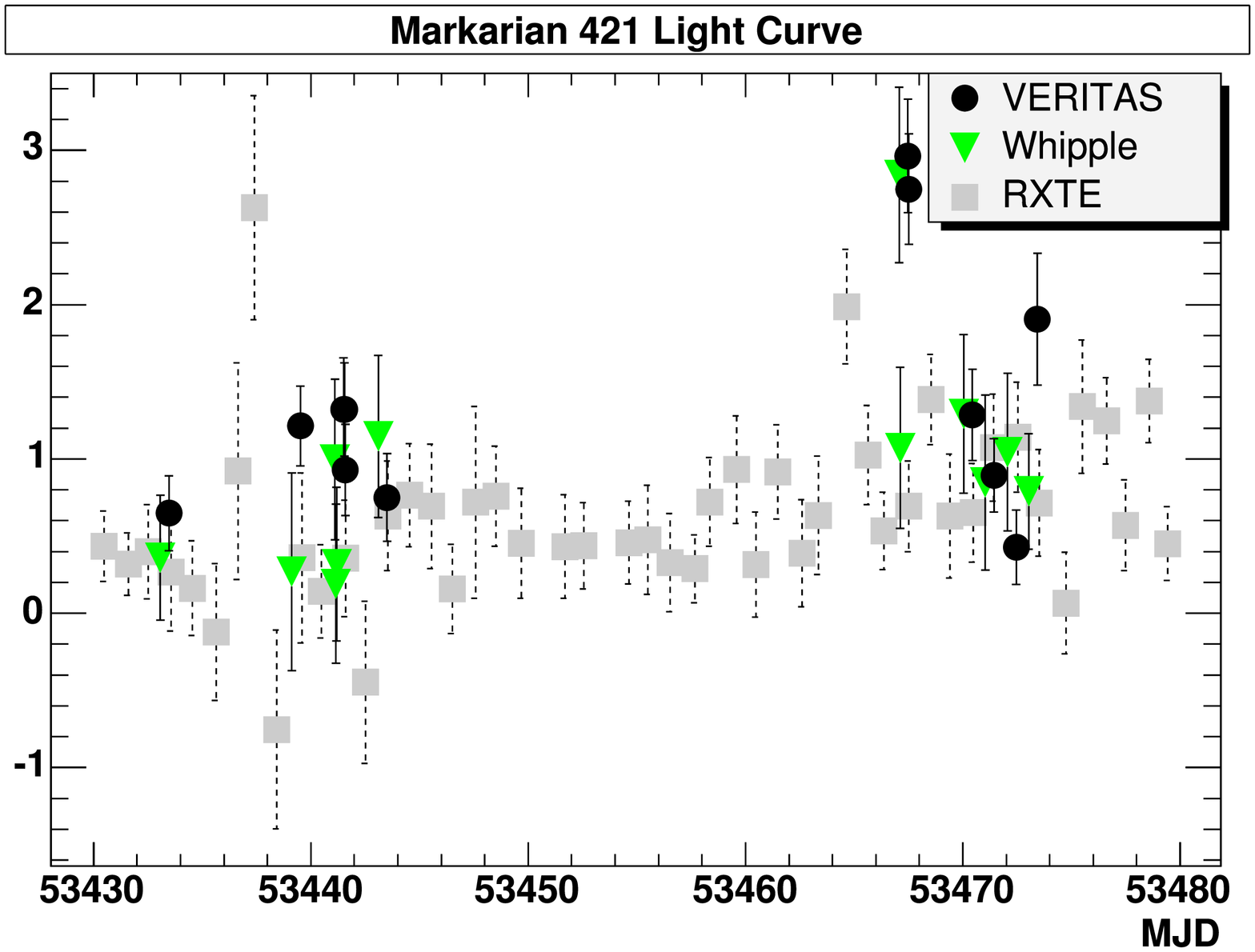}
\caption{\label{lightcurves} \emph{Left:} Light curve for Markarian 501. \emph{Right:} Light Curve for Markarian 421. VERITAS/Whipple points are in $\gamma$/min and RXTE ASM points are in counts/sec.}
\end{center}
\end{figure}

Charges are derived from the FADC traces by integrating over a window
that is centered on the FADC trace and set in a two pass optimised
filter \cite{holder} to maximise the signal to noise ratio (Figure
\ref{trace}). After identification of dead channels, subtraction of
pedestals and relative gain correction, images are prepared for
parameterisation using picture/boundary cleaning \cite{fegan} with
cuts on the charge of 4.0 (picture) and 2.0 (boundary) times the pedestal
RMS. Standard image parameterisation \cite{fegan} is implemented and
cuts, optimised on observations taken on the Crab Nebula, are
applied. Distributions of the image parameter alpha \cite{fegan}, are
shown in Figure \ref{alphaplots} for Markarian 421 and Markarian
501. For point sources located at the center of the field of view, the
alpha distribution should peak at low values. In each case, the alpha
distribution (after application of cuts \cite{maier}) for the ON
source field is shown with the OFF field overlayed. Two dimensional
significance maps, smoothed with a simple boxcar algorithm, are
presented in Figure \ref{skymaps}. Observations of Markarian 421
yielded a $12.5 \sigma$ detection over 13.1 hours, whereas
observations of Markarian 501 yielded a $7.1 \sigma $ detection over
6.7 hours. Finally, light curves are shown in Figure \ref{lightcurves}
with Whipple 10m data points for comparison. The Whipple 10m
data were taken within, at most, one hour of the VERITAS Telescope 1
data. Also shown are ASM/RXTE\cite{rxte} quicklook daily averages.

\section{Conclusions}

VERITAS Telescope 1 operated at a temporary location at the basecamp of the Whipple
Observatory from January to May 2005. During this engineering period,
the Telescope successfully detected the Crab Nebula, Markarian 421 and
Markarian 501. The telescope performed well without light cones in an
area with significant background light, at a comparatively low
elevation and with analysis that was not fully optimised. Despite
these factors and the short period of operation, Telescope 1 met all
(and exceeded many) expectations and proved to be a superior
instrument to the Whipple 10m Telescope.

The VERITAS Collaboration is currently constructing a second identical
telescope at the basecamp of the Whipple Observatory. As well testing
the array trigger and improving background rejection, this
configuration will significantly reduce the energy threshold by
suppressing triggers on local muons.

\section{Acknowledgments}

%The technical assistance of E. Roache, E. Little and R. Pepe is
%acknowledged. 
Funding from The Irish Research Council for Science,
Engineering and Technology(IRCSET): funded by the National Development
Plan. VERITAS is funded by the DOE, NSF, Smithsonian, SFI, PPARC and
NSERC. %The ASM quicklook results were provided by the ASM/RXTE team
%(http://xte.mit.edu).

%VERITAS is funded in the US by the DOE, NSF
%and Smithsonian, in Ireland by SFI, in Canada by NSREC, and in the UK
%by PPARC. The ASM quicklook results were provided by the ASM/RXTE team
%(http://xte.mit.edu).


\begin{thebibliography}{99}


\bibitem{bloommarscher}
Bloom, S.D. and Marscher, A.P. 1996, ApJ 461, 657+

\bibitem{bradburyetal}
Bradbury, S. M. , et al. Proc. 26th ICRC (Salt Lake)

\bibitem{buckleyetal}
Buckley, J. et al. Proc. 28th ICRC (Tsukuba 2003)

\bibitem{cogan}
Cogan, P. Ap\&SS 2005 297 275-281

\bibitem{daviescotton} 
Davies, J and Cotton, E.  Journal of Solar Energy 1 16 1957

\bibitem{fegan}
Fegan, D.J., J. Phys. G: Nucl. Part. Phys. 23 (1997) 1013-1060

\bibitem{hall}
Hall J. et al. Proc. 28th ICRC, (Tsukuba 2003)

%\bibitem{halzen}
%Halzen, F. and Hooper, D.  ASTRO-PH 0502449

%\bibitem{helene}
%Helene, O., 1983, NIM, 212, 319

\bibitem{holder}
Holder, J. et al., Proc. 29th ICRC, (Pune 2005)

%\bibitem{horan}
%Horan, D. et al., 2004, ApJ 603 51-61

%\bibitem{kraw}
%Krawczynski, H. et al., 2004 ApJ 601 151-164

%\bibitem{konigl}
%Konigl, A. ApJ 1981 243, 700

%\bibitem{lebohec}
%Le Bohec, S., Holder J., 2003 Astropart.Phys. 19  221-233

\bibitem{maier}
Maier, G. et al. Proc. 29th ICRC, (Pune 2005)

\bibitem{mannheim}
Mannheim, K.: 1993, A\&A 269, 67

%\bibitem{punch}
%Punch, M. et al: Proc. 22nd ICRC, (Dublin 1991)

\bibitem{rxte}
Quick-look results provided by the ASM/RXTE team.%http://xte.mit.edu/asmlc/srcs

\bibitem{schroedter}
Schroedter, M. , et al. 

\bibitem{urryandpadovani}
Urry, C.M. and Padovani, P. IAU Symp 175: Extragalactic Radio Sources, pp379+








\end{thebibliography}
\end{document}